\documentclass[aps,superscriptaddress,altaffilletter,tightenlines]{revtex4}

\begin{document}
\title{EXACT SELF-INTERACTING SCALAR FIELD COSMOLOGIES}

\author{Luis P. Chimento} 
      
\affiliation {Departamento de F\'{\i}sica, Facultad de Ciencias Exactas y Naturales, \\
Universidad de Buenos Aires, Ciudad  Universitaria,  Pabell\'{o}n  I, \\
1428 Buenos Aires, Argentina.}

\author{Adriana E. Cossarini}
\affiliation {Departamento de F\'{\i}sica, Facultad de Ciencias Exactas y Naturales, \\
Universidad de Buenos Aires, Ciudad  Universitaria,  Pabell\'{o}n  I, \\
1428 Buenos Aires, Argentina.}

\author{Alejandro S. Jakubi }
\affiliation {Departamento de F\'{\i}sica, Facultad de Ciencias Exactas y Naturales, \\
Universidad de Buenos Aires, Ciudad  Universitaria,  Pabell\'{o}n  I, \\
1428 Buenos Aires, Argentina.}

\begin{abstract}
We solve isotropic, homogeneous cosmological models containing a
self-interacting scalar field. Calculations are performed in four and
two-dimensional spacetimes. We find several exact solutions that have an
inflationary regime or has a final Friedmann stage. Also their
asymptotically stability is studied.
\end{abstract}
\maketitle

\section{Introduction}

A self-interacting scalar field has been introduced in cosmological models
as a matter source to the Einstein equations because, when dominated by the
potential energy, it violates the strong energy condition and drives the
universe into an inflationary period \cite{Kolb}. 
Currently, there is no underlying principle that uniquely specifies the
potential for the scalar field and many proposals have been considered. Some
were based in new particle physics and gravitational theories, while others
were postulated ad-hoc to obtain the desired evolution \cite{Ell}. Also, a
formalism has been proposed to reconstruct the potential from knowledge of
tensor gravitational spectrum or the scalar density fluctuation spectrum
\cite{Cop}

Due to the non-linearity of the system of differential equations for the scalar
and gravitational fields very little is known yet about exact solutions of
these cosmological models. In section 2 of this paper we show a procedure to
reduce to quadratures the Einstein and scalar field equations in a
Robertson-Walker metric. Our procedure allows for an arbitrary potential and 
we show as an example new exact solutions. We study  their asymptotic
stability by means of the method of Lyapunov \cite{Kra}. No a priori assumption
like a slowly varying field is required to perform the calculations, and we
check the validity of this assumption. Two-dimensional spacetimes are nowadays
very useful for testing ideas on quantum gravity. For this reason, in section 3
we apply our procedure also in this case. The conclusions are stated in section
4.

\section{The Einstein-Scalar Field Equations}

We wish to investigate the evolution of a universe with a scalar field  $
\phi $ which has a self-interaction potential $V(\phi )$ and is minimally
coupled to gravity 

\begin{equation}
\label{1}
\nabla^\mu\nabla_\mu\phi +{\frac{dV}{d\phi }}=0
\end{equation}
where $\nabla^\mu$ is the covariant derivative. Thus, we must solve Eq. \ref{1} together with the Einstein equations

\begin{equation}
\label{3}
R_{ik}-\frac 12g_{ik}R=T_{ik}^\phi 
\end{equation}
We are using units such that $c=8\pi G=1$ and 

\begin{equation}
\label{4a}
T_{ik}^\phi =\phi _{;i}\phi _{;k}-g_{ik}\left( {\frac 12}\phi
_{;m}\phi ^{;m}-V(\phi )\right) 
\end{equation}

\noindent is the stress-energy tensor of the field. In a spatially flat
Robertson-Walker metric

\begin{equation}
\label{5}
ds^2=dt^2-{a^2(t)}\left(dx_1^2+dx_2^2+ dx_3 ^2\right)
\end{equation}

\noindent with scale factor $a(t)$, Eqs. \ref{1} and \ref{3} become

\begin{equation}
\label{6}
\ddot \phi +3H\dot \phi +{\frac{dV}{d\phi }}=0
\end{equation}
\begin{equation}
\label{7}
3H^2={\frac 12\dot \phi }^2+V(\phi )
\end{equation}

\noindent where the dot means $d/dt$, $H=\dot a/a$, and $\phi =\phi (t)$.
It becomes convenient to use the scale factor as the independent variable
and write the potential in the following form:

\begin{equation}
\label{10} 
V[\phi (a)] = {\frac{F(a)}{a^{6}}} 
\end{equation}

\noindent with a suitable function $F(a)$. Thus we obtain a first integral
of Eq. \ref{6}

\begin{equation}
\label{11} 
{\frac 12\dot \phi }^2+V(\phi )-{\frac 6{a^6}\ \int }da{\frac Fa}=
{\ \frac C{a^6}} 
\end{equation}

\noindent where $C$ is an arbitrary integration constant. Then, using Eqs. 
\ref{7} and \ref{11}, we have reduced the problem to quadratures:

\begin{equation}
\label{12}
\Delta t={\sqrt{3}\int \frac{da}a}\left[ {\frac 6{a^6}\ \int }da{
\frac Fa+\frac C{a^6}}\right] ^{-1/2}
\end{equation}

\begin{equation}
\label{13}
\Delta \phi ={\sqrt{6}\int \frac{da}a\left[ \frac{-F+6\int daF/a+C
}{6\int daF/a+C}\right] }^{1/2}
\end{equation}

\noindent where $\Delta t\equiv t-t_0$, $\Delta \phi \equiv \phi -\phi _0$
and $t_0$, $\phi _0$ are arbitrary integration constants.

\subsection{ Example}

As an simple example of our procedure let us consider the function

\begin{equation}
\label{20}
F(a)=B a^s\left( b+a^s\right) ^n
\end{equation}

\noindent  This is interesting because it yields new exact solutions that
exhibit stable exponential or Friedmann behavior. In Eq. \ref{20}, $B>0$, $b> 0$,
$s$ and $n$ are constants and we require that $s(n+1)=6$.  Then, inserting Eq.
\ref{20} in Eqs. \ref{13} and
\ref{10} and taking for simplicity $C=0$ we get

\begin{equation}
\label{22}
V(\phi )=B\left[ \cosh \left( {\frac s{2\sqrt{6}}}\Delta \phi \right) \right]
^{2n} 
\end{equation}

\noindent This potential has a nonvanishing minimum at $\Delta \phi =0$ for $
s>0$, which is equivalent to a  effective cosmological constant.
When $s<0$ , the origin becomes a maximum, and the potential vanishes
exponentially for large $\phi $. We can evaluate Eq. \ref{12}  for some values
of $s$, for instance:

\begin{equation}
\label{26}
\Delta t={\sqrt{3\over B}}\left[ {\rm arcsinh} \left({\frac a{\sqrt{b}}}\right)
-{\frac a{(b+a^2)^{1/2}} }\right] ,\qquad s=2
\end{equation}

\begin{equation}
\label{27}
a=\left\{b\left[\exp \left(\sqrt{3B}\Delta
t\right)-1\right]\right\}^{1/3} , \qquad\qquad\qquad s=3  
\end{equation}

\noindent For $s>0$, the evolution begins from a singularity as ${\Delta
t}^{1/3}$ and is asymptotically de~Sitter with 
$\Delta\phi\rightarrow 0$ for $t\rightarrow\infty$. On the other hand, for
$s<0$ the evolution has a deflationary behaviour from a de Sitter era in the
far past to a Friedmann   behavior ${\Delta t}^{1/3}$ when
$t\rightarrow\infty$.

\subsection{Slow-Roll Approximation}

A common framework to solve Eqs. \ref{6}, \ref{7} in discussions of
inflation is the ''slow-roll'' approximation \cite{Kolb}. To investigate its
limitations  we  calculate the two slow-roll parameters \cite{Cop93}

\begin{equation}
\label{16a} 
\epsilon \equiv {\frac{{\dot \phi }^2/2}{V+{\dot \phi }^2/2}}=1-{
\ \frac F{C+6\int daF/a}} 
\end{equation}

\begin{equation}
\label{16b} 
\eta \equiv {\frac{\ddot \phi }{H\dot \phi }}={\frac{
12F-aF^{\prime }-36\int daF/a-6C}{2\left( C+6\int daF/a+F\right) }} 
\end{equation}

The parameter $\epsilon $ measures the relative contribution of the field's
kinetic energy to its total energy density and $\eta $ measures the ratio of
the field's acceleration relative to the friction term. The slow-roll
approximation is valid when $\mid \epsilon \mid \ll 1$ and $\mid \eta \mid
\ll 1.$ Using Eqs. \ref{16a} and \ref{16b}, we find that they impose a
constrain on the form of the potential and the value of the initial
conditions. For instance, in our example they imply that $b/a^s\ll 1$. Thus, it
is violated for short times if $s>0$ or long times if $s<0$.

\subsection{ Stability of the Solutions}

For models (like our example for $s>0$) such that $V(\phi )$ has a local
minimum at $\phi _m$ and $V(\phi _m)\ge 0$, we can study the stability of
solutions with asymptotic behavior $\phi (t)\rightarrow \phi _m$. First we note
that the evolution a(t) is monotonic. Then, differentiating Eq. \ref{7} and
using Eq. \ref{6}, we find $2\dot H=-\dot \phi ^2<0$ so that these models have
a sub-inflationary behavior. As $H\ge 0$ and $dH/da\le 0$ in a neighborhood of
$\phi _m$, $ H(\phi ,a)$ is a Lyapunov function for the system Eqs. \ref{6} and
\ref{7}, and any solution such that $\phi \rightarrow \phi _m$ for
$a\rightarrow \infty $ (equivalently $t\rightarrow \infty $), is asymptotically
stable.

\section{Two-Dimensional Spacetime}

In recent years, there have been a  number  of  investigations
into the structure of relativistic gravitational theories in two 
spacetime  dimensions  \cite{Pol} ${}^-$\cite{Man}  mainly  because  they
reduce   the complexity of four-dimensional general relativity and  constitute
useful testing grounds for ideas on quantum gravity.  Among  them, the so
called "$R = T$" theory \cite{Wit} has attracted  some  interest due to the
fact that it  has  many  classical  aspects  which  are similar to general
relativity.

In  this  paper,  we  confine  our   attention   to   classical
cosmological properties of an two-dimensional universe based  on
the gravitational field equations:

\begin{equation}
R = 8\pi G T \qquad\qquad \nabla ^{\mu }T_{\mu \nu }= 0
\label{1.1}
\end{equation}

\noindent where $R$ is the  curvature  scalar  and $T$  is  the  trace  of  the
energy-momentum tensor.  We consider the two-dimensional  form  of
the Robertson-Walker metric filled with a minimally coupled scalar
field $\phi $ with a self interacting potential $V(\phi )$  which  obeys  
the Klein-Gordon Eq. \ref{1}.
 Inserting $T = 2 V(\phi )$ and $R =- 2 \ddot a/a$   in Eq. \ref{1.1},  the
system of Eqs. \ref{1}  and  \ref{1.1} become:

\begin{equation}
\ddot \phi + H \dot \phi + {dV\over d\phi } = 0
\label {2.7}
\end{equation}

\begin{equation}
\ddot a = -a V(\phi )
\label{2.8}
\end{equation}

\noindent Following the steps of the previous section we write the potential in
the form

\begin{equation}
V[\phi (a)] = {F(a)\over a^{2}}
\label{2.9}
\end{equation}

\noindent and we  obtain  two  first  integrals of Eqs. \ref{1.1} and \ref{1}:

\begin{equation}
{1\over 2}\dot \phi^2 + V(\phi ) - {2\over a^{2}} \int  da {F\over a} =
{C_{1}\over a^{2}}
\label{2.10}
\end{equation}

\begin{equation}
{\dot a^{2}\over 2} + \int  da {F\over a} = C_{2}
\label{2.11}
\end{equation}

\noindent where $C_{1}$ and $C_{2}$ are  arbitrary  integration  constants.
Then  the problem have been reduced to quadratures:

\begin{equation}
\Delta t = \int  da \left[ 2 C_{2} - 2 \int  da {F\over a} 
\right]^{-1/2}
\label{2.12}
\end{equation}

\begin{equation}
\Delta \phi  = \int  {da\over a} \left[ {{C_{1} - F + 2\int da F/a}
 \over {C_{2} - \int da F/a}}\right] ^{1/2} 
\label {2.13}
\end{equation}

Eqs. \ref{2.12} and \ref{2.13}  show  two  important   differences  with
respect to their four dimensional counterparts. The first  one  is
the appearance of  two different independent constants $C_{1}$  and $C_{2}$
while in four dimensions we have the  constrain $C_{1} = C_{2}$.  The
second one is the sign of the terms proportional to $\int da \,F/a$

\subsection{Example}

In the case of Eq.  \ref{20}  we find an exact solution taking $s(n+1) = 2$, 
$s>0$  and $C_{1}=C_{2}=0$ in Eqs. \ref{2.12} and \ref{2.13}. 
  Taking into account that the Eqs. \ref{2.10} and \ref{2.11} lead to  the
constrains $B<0$, $b<0,$ the potential results

\begin{equation}
V(\phi ) = B \sin ^{2n}\left( {s\over {2\sqrt 2}} \Delta \phi \right)
\label{3.21}
\end{equation}

 Contrary to the results obtained in the four dimensional case,  we
obtained a negative periodic potential  which  oscillates  between
the values $V_{\min }= B<0$ and $V_{\max }=0$. This is related to
the change in sign we  pointed  out  in  the  previous subsection. 
 We can evaluate Eq. \ref{2.12} for some particular values of $s$:

\begin{equation}
a(t) = b\left[\exp \left(\sqrt {-B}\Delta t\right) - 1\right],
\quad\qquad\qquad\qquad\qquad s=1
\label{3.24a}
\end{equation}

\begin{equation}
\Delta t = {3\over \sqrt{-B}} \left[
{\rm arccosh}\left({a^{1/3}\over \sqrt b}\right) - \left({a^{2/3}\over {b + a^{2/3}}}
\right)^{1/2}\right],
\qquad s=2/3
\label{3.24b}
\end{equation}

Thus, in two dimensions the evolution behaves like $\Delta t$ near the
singularity and, notwithstanding the fact that $V(\phi)<0$, it has an
asymptoticaly de Sitter behavior for $t \rightarrow \infty$ as in the
four-dimensional case. 

\section{Conclusions}

We present a procedure which reduces to quadratures the gravitational field
equations with a classical self-interacting scalar field as a matter source in
a Robertson-Walker spacetime. The freedom to choose the potential of the scalar
field is expressed in terms of the function $F(a)$, which we can select at
will. We take a simple example of this function and we show that it yields new
stable exact solutions in the four-dimensional spacetime. We analyse the
restrictions imposed by the "slow-roll" approximation and we verify in our
example that they may very easily violated. In the two-dimensional case, we
show that the same function $F(a)$ leads to a negative oscillating potential.
The evolution changes its behavior near the singularity but remains the same in
the far future.

In a future paper we will extend our procedure to more general models with 
curvature term, a cosmological constant and a perfect fluid source.


\section{References}

\begin{thebibliography}{9}
\leftmargin 2.5em

\bibitem{Kolb}
E.~W.~Kolb and M.~S.~Turner
{\it The Early Universe \/} (Addison-Wesley, New York, 1990).

\bibitem{Ell}
G.~F.~R.~Ellis and M.~S.~Madsen
{\it Class. Quantum Grav. \/}  {\bf 8} (1991) 667.

\bibitem{Cop}
E.~J.~Copeland, E.~W.~Kolb, A.~R.~Liddle and J.~E.~Lidsey
{\it Phys. Rev. Lett. \/} {\bf 71} (1993) 219.

\bibitem{Kra} 
N.~N.~ Krasovskii 
{\it Stability of Motion \/} (Stanford University Press, Stanford, 1963).

\bibitem{Cop93}
E.~J.~Copeland, E.~W.~Kolb, A.~R.~Liddle and J.~E.~Lidsey
Fermilab preprint FERMILAB-PUB-93/029-A.

\bibitem{Pol}
A.~M.~Polyakov
{\it Mod. Phys. Lett. \/} {\bf A2} (1987) 893.

\bibitem{Chi}
L.~P.~Chimento and A.~E.~Cossarini
{\it Nucl. Phys.} {\bf B373} (1992) 438.

\bibitem{Man}
R.~B.~Mann 
{\it Fond. Phys. Lett. \/} {\bf 4} (1991) 425.

\bibitem{Wit}
E. Witten
{\it Phys. Rev. \/} D {\bf 44} (1991) 314.

\end{thebibliography}
\end{document}